\documentclass[useAMS,usenatbib,referee]{mn2e}

\usepackage{epsfig}
\usepackage{amsmath}
\usepackage{amssymb}
\usepackage{natbib}

\usepackage{epstopdf}

\def \mnras {MNRAS}
\def \apj {ApJ}

\def \apjl {ApJL}
\def \aap {A\&A}

\def \pasj {PASJ}

\title[XMM-Newton observations of IGR J17091--3624]{The enigmatic black-hole candidate and X-ray transient IGR J17091--3624 in its quiescent state as seen with {\itshape XMM-Newton}}
\author[Wijnands et al.]
{R. Wijnands$^{1}$\thanks{e-mail: r.a.d.wijnands@uva.nl},  
Y.J. Yang$^{1}$,
D. Altamirano$^1$ \\
$^{1}$Astronomical Institute "Anton Pannekoek", 
University of Amsterdam, 
Postbus 94249, 1090 GE Amsterdam, the Netherlands
}

\begin{document}


\pagerange{\pageref{firstpage}--\pageref{lastpage}} \pubyear{0000}

\maketitle

\label{firstpage}

\begin{abstract} We report on two short {\it XMM-Newton} observations
performed in August 2006 and February 2007 during the quiescence state
of the enigmatic black hole candidate system IGR J17091--3624. During
both observations the source was clearly detected. Although the errors
on the estimated fluxes are large, the source appears to be brighter
by several tens of percents during the February 2007 observation
compared to the August 2006 observation.  During both observations the
2--10 keV luminosity of the source was close to $\sim10^{33}$ erg
s$^{-1}$ for an assumed distance of 10 kpc. However, we note that the
distance to this source is not well constrained and it has been
suggested it might be as far as 35 kpc which would result in an order
of magnitude higher luminosities.  If the empirically found relation
between the orbital period and the quiescence luminosity of black hole
transients is also valid for IGR J17091--3624, then we can estimate an
orbital period of $>$100 hours ($>$4 days) for a distance of 10 kpc
but it could be as large as tens of days if the source is truly much
further away. Such a large orbital period would be similar to GRS
1915+105 which has an orbital period of $\sim$34 days. Orbital periods
this large could possibly be connected to the fact that both sources
exhibit the same very violent and extreme rapid X-ray variability
which has so far not yet been seen from any other black hole
system. Alternatively the orbital period of IGR J17091--3624 might be
more in line with the other systems ($<$100 hours) but we happened to
have observed the source in an episode of elevated accretion which was
significantly higher than its true quiescent accretion rate. In that
case, the absence or presence of extreme short-term variability
properties as is seen for IGR J17091--3624 and GRS 1915+105 is not
related to the orbital periods of these black hole systems.
 \end{abstract}

\begin{keywords}
X-rays: binaries - binaries: close - 
stars: individual (IGR J17091--3624) - black hole physics
\end{keywords}

\section{Introduction} 

Accretion neutron stars and black holes in X-ray transients are
systems which are typically found in a dormant, quiescent state during
which they cannot be detected in X-rays or only at very low
luminosities. In such quiescent states no or hardly any accretion
occurs onto the compact stars. However, occasionally they go in
outbursts during which their X-ray luminosities increase by several
orders of magnitude and such systems become visible as X-ray
binaries. This huge increase in brightness is caused, presumably, by a
similarly large temporary increase in accretion rate onto the
accretors.  Typically X-ray binaries are divided into two general
sub-classes: (i) low-mass X-ray binaries in which the donor mass is
lower than the mass of the accretor and matter transfer occurs because
the donor fills its Roche-lobe, and (ii) high-mass X-ray binaries in
which the donor mass is higher than the mass of the accretor, and
matter transfer occurs via the strong stellar wind of the companion or
due to a Be excretion disk. Often, when a new transient X-ray binary
is discovered it is not directly clear what kind of accretor and donor
it has and the system is classified based on its similarities with
other, known types of systems.

When in outburst the X-ray transients can easily be studied due to the
large number of photons observed from those systems. However, only a
few X-ray satellites have the sensitivity to detect the very faint
X-ray emission during the quiescent state of these systems. {\it
Chandra} and {\it XMM-Newton} have proven crucial to make significant
progress in detecting many systems in quiescence and have allowed the
study of several of the brightest ones in great detail. One of the
main findings is that black hole transients are systematically (albeit
not always) significantly fainter in quiescence than the neutron stars
systems, especially when comparing them at the same orbital period
\citep{1997ApJ...478L..79N,2001ApJ...553L..47G,2002ApJ...570..277K}. This
difference has been used as evidence that black holes have event
horizons while neutron stars have solid surfaces
\citep[e.g.,][]{1997ApJ...478L..79N,2001ApJ...553L..47G,2002nmgm.meet..405N,2004ApJ...615..402M,2008NewAR..51..733N}.
This hypothesis is consistent with the observed fact that in many
neutron star systems a soft thermal component (with black body
temperatures below 0.2-0.3 keV) has been observed \citep[][and
reference to those
papers]{1996PASJ...48L..27A,1996PASJ...48..257A,1999ApJ...514..945R},
presumably from the neutron star surface. Such soft components have
never been seen in the quiescent state of black hole transients
\citep[][]{2004ApJ...615..402M} which would be consistent with the
absence of a solid surface.  The quiescent X-ray spectra of black hole
transients can typically be described with a single power-law model
with a photon index of around 2 \citep[e.g., ][]{2002ApJ...570..277K}
albeit usually with large error bars because of the often very low
number of photons detected from those systems.

The origin of the X-ray emission of quiescent black hole transients is
not very well understood. The most frequently used model assumes that
there is still residual accretion occurring onto the black hole even
in quiescence, but that the accretion occurs through a so-called ADAF
(advection dominated accretion flow; see \citet{1997ApJ...478L..79N}
and references to that paper) type of flow which is very inefficient
in radiating its energy away and most of the energy from the
in-falling matter is advected beyond the event horizon. Therefore, the
black hole X-ray transients can be very faint. If indeed this is the
correct explanation for the quiescent emission of black hole
transients, the X-ray emission of these systems is predicted to be
correlated with the orbital period of the binary: the luminosity
should be larger for larger orbital periods
\citep[][]{1999ApJ...520..276M}. Such a trend, albeit with significant
scatter, has indeed been observed
\citep[][]{2001ApJ...553L..47G,2002ApJ...570..277K,2004ApJ...615..402M}
giving weight to the ADAF interpretation for the quiescent X-ray
emission of black holes.

To further test and constrain the ADAF interpretation, more black
holes in quiescence must be detected and studied in detail. An
excellent candidate would be IGR J17091--3624. This source was
discovered in April 2003 using {\it INTEGRAL}
\citep{2003ATel..149....1K}.  Another outburst was seen in 2007
\citep{2007ATel.1140....1K,2009ApJ...690.1621C} and archival studies
showed that the source had been previously active during several other
occasions \citep{2003ATel..150....1R,2003ATel..160....1I}. Based on
its outbursts properties (i.e., its X-ray spectral and timing
behavior) the source was suggested \citep{2006ApJ...643..376C} to
harbor a black hole as the accretor, although a neutron star could not
be excluded.  In February 2011, the source was detected again in
outburst \citep[using {\it
Swift};][]{2011ATel.3144....1K,2011ATel.3148....1K} but this time the
source stayed on until at least the time of writing this letter ({\it
Swift} observations performed on 26 and 31 January 2012 showed the
source still to be active). The 2011 outburst of the source has caused
quite some excitement because in the X-rays it suddenly displayed very
unusual X-ray variability which so far had only been seen in the
enigmatic very bright black hole X-ray transient GRS 1915+105
\citep[e.g., see][for details about its variability and its
comparison with GRS
1915+105]{2011ApJ...742L..17A,2012arXiv1201.2106A,2011arXiv1105.4694P}.
 \citet{2009ApJ...690.1621C} reported on two {\it XMM-Newton}
observations taken in 2006 and 2007 during which IGR J17091--3624 was
in its quiescent state.  They claim that the source could not be
detected during these observations, however, we have re-examined those
{\it XMM-Newton} observations and we clearly detect a source in both
observations. In this letter we report on those observations and
discuss how the source fits in within the general picture of quiescent
black hole transients.

\section{Observations, analysis and results}

{\it XMM-Newton} observed the field containing IGR J17091--3624 on 25
August, 2006, and 19 February 2007 (see Tab.\ref{log}; see also
\citet{2009ApJ...690.1621C}). For all EPIC camera's the medium filter
was used. We do not analyze the RGS data in this letter since the
source was too faint (see below) to result in any significant flux
from the source using this instrument. The data were analyzed using
SAS version 11 and following the standard analysis
threads\footnote{http://xmm.esac.esa.int/sas/current/documentation/threads/}.
To apply the most up-to-date calibration we reprocessed the original
data files using the programs {\tt epproc} and {\tt emproc}. We
searched for the presence of background flares in the EPIC data using
only the data above 10 keV and found none during the 2006 observation
but a small flare occurred at the end of the 2007 observation. This
flare was removed from the data; Table~\ref{log} lists the resulting
exposure times. 

\begin{figure}
 \begin{center}
\includegraphics[width=16cm]{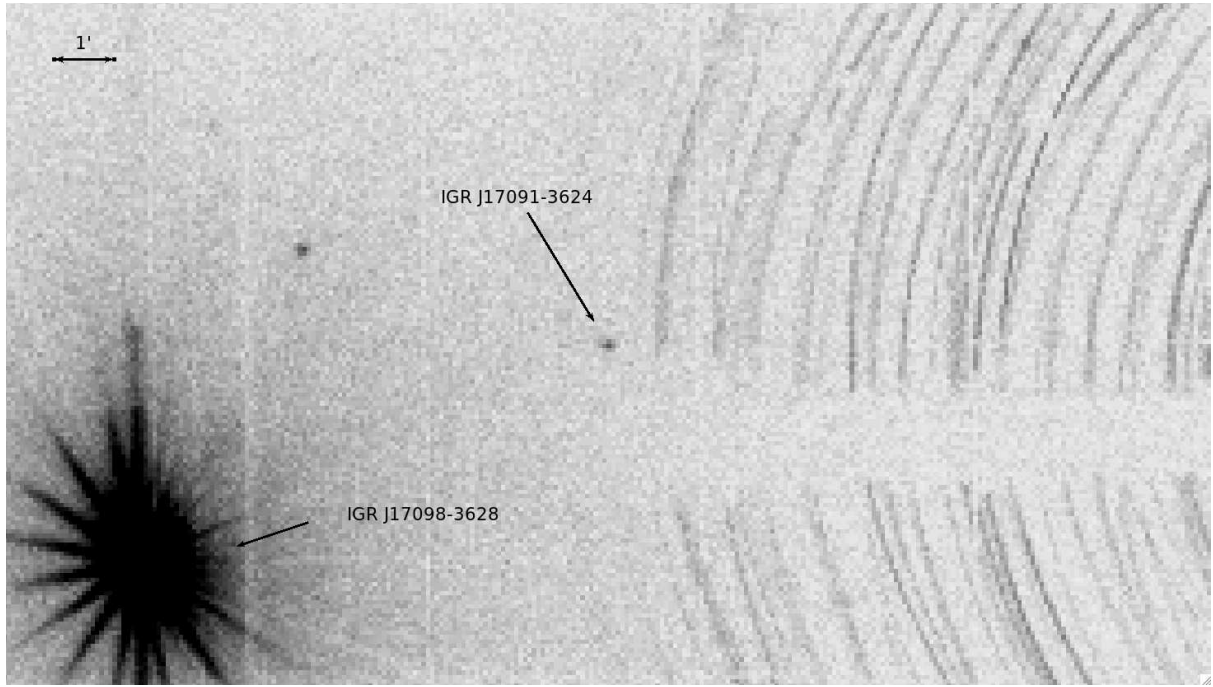}
    \end{center} \caption[]{{\it XMM-Newton} images (both observations and
    all EPIC detectors
    combined) of IGR J17091--3624. The source as well as the close-by
    active transient
    IGR J17098--3628 are indicated by arrows.  The arcs at the
    right are due to
    the bright persistent neutron-star X-ray binary GX 349+2 which is
    outside the field of view.}
 \label{fig1}
\end{figure}

Figure~\ref{fig1} shows the combined 0.5--10 keV image of all data
(including both observations and all three EPIC instruments; the
contaminating arcs are due to the bright neutron-star X-ray binary GX
349+2 which is outside the field-of-view of the telescope).  Contrary
to the findings by \citet{2009ApJ...690.1621C}, we clearly detect a
source close to the radio position reported for IGR J17091--3624
\citep[][]{2011ATel.3167....1C}. The source was most clearly visible
in the pn images so we used the pn data and the tool {\tt
edetect\_chain} to extract the source position (using only the 0.5--10
keV data). The best source position was obtained from the 2007
observation: right ascension = $17^h ~ 09^m ~ 07.674^s$ and
declination = $-36^\circ ~24^{'} ~ 25.3^{''}$ (epoch J2000) with a
statistical error of 0.9$^{''}$ and $\sim2^{''}$ ($1\sigma$) absolute
astrometry error\footnote{The {\it XMM-Newton} calibration documents
can be found at
http://xmm.vilspa.esa.es/external/xmm\_sw\_cal/calib/documentation.shtml}. This
position (as well as the position obtained during the 2006 {\it
XMM-Newton} observation) is fully consistent with the radio position
of the source and it is very likely that the detected source is the
quiescent X-ray counterpart of IGR J17091--3624. The 0.5--10 keV count
rates (using the pn) of the source during the first and second
observations were 0.012$\pm$0.003 counts s$^{-1}$ and 0.020$\pm$0.002
counts s$^{-1}$ respectively. This difference in count rate indicates
that during the second observation the source was slightly brighter
than during the first observation.  We searched for variability in the
light curves during each observations but the statistics were very
limited inhibiting any conclusion on this.

\begin{figure}
 \begin{center}
\includegraphics[angle=-90,width=16cm]{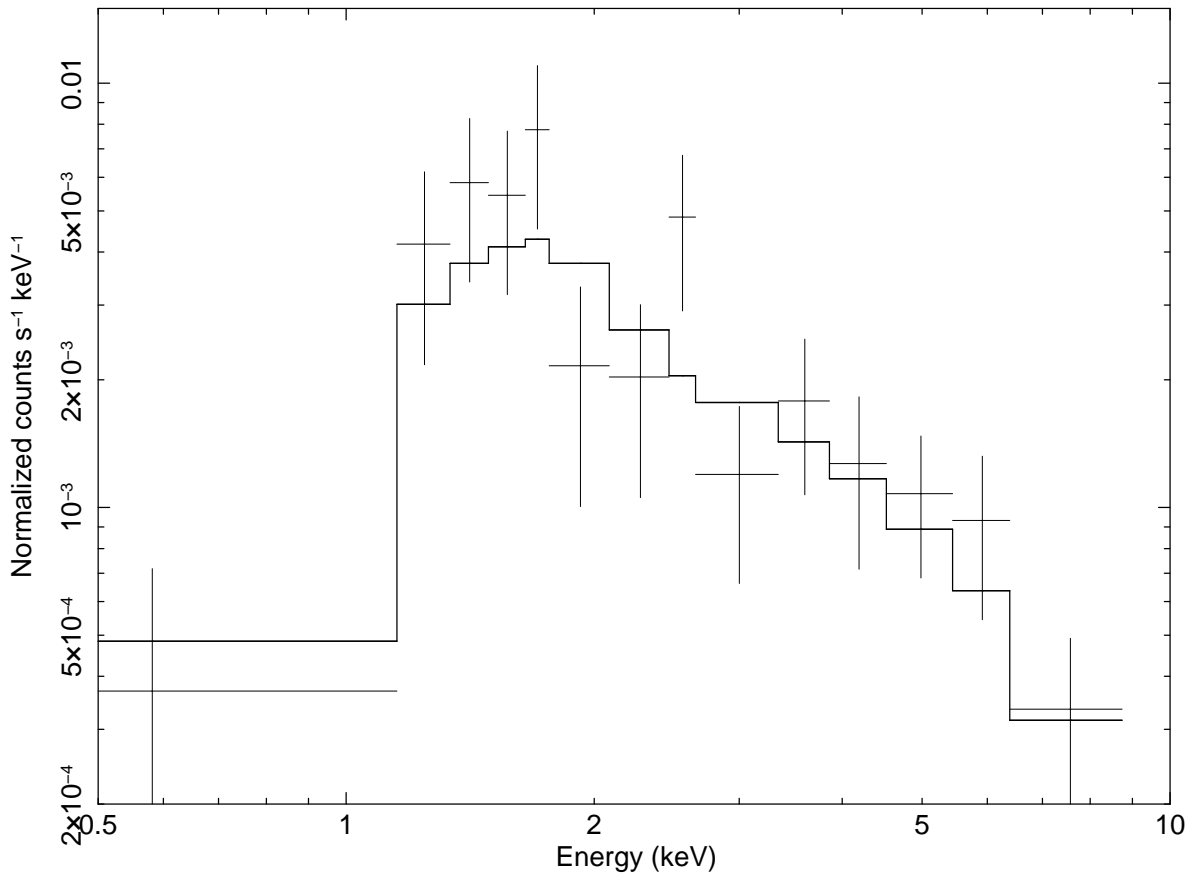}
    \end{center} \caption[]{{\it XMM-Newton} pn spectrum of the 2007 observation. The solid line through the data points is the best fitted absorbed power-law model.}
 \label{fig2}
\end{figure}

To extract the source spectrum we used a circle of 10$^{''}$ on the
source position. To estimate the background we used a circle of
25$^{''}$ on a region of the CCD with was free of sources and also
free of the contaminating arcs in the image. We only report here on
the pn spectra because very limited number of photons were recorded
using the MOS camera's (e.g., in the MOS2 data of the first
observation the source could not conclusively be detected). The
spectra were extracted using the tool {\tt especget}. The spectra were
rebinned to have at least 10 photons per bin.  The spectral data were
analyzed using Xspec version 12.7.0. The pn spectrum obtained during
the 2007 observation is shown in Figure~\ref{fig2}. The quality of the
data is insufficient to perform a detailed spectral analysis and we
only tried to fit the data with single component models. However,
since quiescent spectra of black hole transients are typically fitted
with a simple power-law model we focus only on that model but we note
that a black body or a multi-color disk black body model could equally
well fit the data.

We fitted the spectra with an absorbed power-law model in which the
column density was fixed to the outburst value of $1.1 \times 10^{22}$
cm$^{-2}$ \citep[][]{2011A&A...533L...4R} and the photon index was
tied between the observations. The limited quality of the data does
not allow to constrain those parameters independently. The
normalization was left free in order to investigate possible
variability between the two observations. This model could fit the
data adequately with a reduced $\chi^2$ of 0.7 with 13 degrees of
freedom. The obtained photon index was 1.6$\pm$0.5 and the obtained
fluxes are reported in Table~\ref{log}. The fluxes also show (similar
to the count rates) that the source likely was fainter during the 2006
observation compared to the 2007 observation although the errors on
the fluxes are large and formally the fluxes are consistent with each
other.

\begin{table}
 \caption{The log of the {\it XMM-Newton} observations\label{log}}
 \begin{tabular}{c|r|c|ccc}
  ObsID      & Date of observation             & Useful exposure$^1$ & Energy range & Observed flux          & Unabsorbed flux \\
             &                                 &   (ksec)              & (keV)            & $\times 10^{-14}$erg s$^{-1}$ cm$^{-2}$ & $\times 10^{-14}$ erg s$^{-1}$ cm$^{-2}$ \\
  \hline
  0406140101 & 25 August 2006   & 5.3  & 2--10    & $8.4\pm2.3 $             & $9.2\pm2.3 $ \\
             &                  &           & 0.5--10 & $9\pm2 $             & $13.6^{+3.4}_{-3.9}$\\
  0406140401 & 19 February 2007 & 11.1  & 2--10   & $11^{+2}_{-3} $    & $12\pm3$\\ 
             &                  &           & 0.5--10 & $12^{+2}_{-3} $    & $18\pm2$\\
  \hline
 \end{tabular}
  $^1$ The exposure time per EPIC camera could be
slightly different due to
variations in the exact moments when the detectors were switched on and
off.
\end{table}

\section{Discussion}\label{sec:discuss}

We report on the X-ray detection (using {\it XMM-Newton}) of the
enigmatic X-ray transient and black hole candidate IGR J17091--3624 in
its quiescent state. The X-ray spectrum of the source could be
described with a simple absorbed power-law model with a photon index
of 1.6$\pm$0.5. This is consistent with what is observed for other
black hole transients in their quiescent states
\citep[e.g.,][]{2002ApJ...570..277K}. The obtained fluxes are listed
in Table~\ref{log}.  Due to lack of constraints on the source
distance, it is not possible to estimate accurately the quiescent
X-ray luminosity of IGR J17091--3624 making it rather difficult to
compare our results to those of other quiescent black hole
systems. Therefore, we will discuss several scenarios based on
different assumptions.

The location of the source in the Galactic bulge might suggest a
distance of only 8 to 10 kpc similar to what is typically assumed for
sources in the bulge. For 10 kpc, the 0.5--10 keV luminosity would be
1-2$ \times 10^{33}$ erg s$^{-1}$ (the range is due to the errors on
the fluxes and the observed flux variation between the two
observations). However, it is possible that the source might be
significantly further away than 10 kpc.  IGR J17091--3624 displayed
atypical behavior during its 2011 outburst. In many ways it resembles
another enigmatic black hole system, namely GRS 1915+105. Both showed
a large variety of variability phenomena in their light curves
although IGR J17091--3624 displayed variability on a time scale of
about a factor of 40-50 faster than seen in GRS 1915+105
\citep{2011ApJ...742L..17A}. This kind of extreme variability behavior
is not understood but for GRS 1915+105 it has been postulated to be
related to the high X-ray luminosities (Eddington to possibly even
super-Eddington luminosities) of the source \citep[see the discussion
and references in][]{2011ApJ...742L..17A}. The X-ray flux of IGR
J17091--3624 during outburst is considerably lower than that of GRS
1915+105 (by a factor of 30). If also IGR J17091--3624 is accreting
near the Eddington limit it must have a black-hole mass of only 3
solar masses for the source to be located within our Galaxy ($<$20
kpc), or the source must be at  $>$35 kpc if it would harbor a black
hole with a mass of $\sim15 M_\odot$ (similar to the black-hole mass
of GRS 1915+105). For such distances, the quiescent 0.5--10 keV X-ray
luminosity of IGR J17091--3624 would be 5-10$\times 10^{33}$ erg
s$^{-1}$ or 1-3$\times10^{34}$ erg s$^{-1}$ for 20 and 35 kpc,
respectively. Irrespective of the source distance, the X-ray
luminosity of this source is among the highest observed for quiescent
black hole transients.

To explain the quiescent X-ray emission of black hole transients
several models have been proposed but the most commonly used model is
that of residual accretion onto the black hole through an ADAF-like
accretion flow. A profound and testable aspect of this model is that
the quiescent luminosity of black hole transients should increase with
the orbital period of the systems
\citep{1999ApJ...520..276M}. Although the observations are still quite
limited, a trend of increasing quiescent luminosities with increasing
orbital periods is indeed suggested by the observations
\citep[][]{2001ApJ...553L..47G,2002ApJ...570..277K,2004ApJ...615..402M}. Using
the data presented by \citet[][see their Figure 4; see also
\citet{2004ApJ...615..402M}]{2011ApJ...734L..17R} we estimate that the
orbital period of the system should be $>100$ hours ($>4$ days) for a
source distance of 10 kpc but it could be up to tens of days for
larger distances.  We cannot offer stronger constraints since no black
holes have been observed in their quiescent states with an orbital
period $>$200 hours and the empirically found relation might break
down for larger orbital periods. If its orbital period indeed turns
out to be this large, the quiescent behavior of IGR J17091--3624
provides strong support for an ADAF-like interpretation of the
quiescent X-ray emission of black hole transients.

It is not truly unexpected that IGR J17091--3624 could have such a
large orbital period because of its similarities during outburst with
GRS 1915+105. This system has a measured orbital period of $\sim$34
days and therefore quite a large accretion disk. Such a large disk
could be related to or maybe even the cause of the very high accretion
luminosities of this system \citep[e.g.,][]{2004MNRAS.349..393D}. If
it is indeed true that the violent variability seen in GRS 1915+105
only can occur at such high luminosities, IGR J17091--3624 must also
be accreting at (near-)Eddington luminosities and possibly a large
disk and a large orbital period might also be required in this system.
In this assumed scenario it remains unclear if IGR J17091--3624
harbors a low-mass black hole at $\sim$20 kpc or a more massive black
hole at $>$35 kpc \citep{2011ApJ...742L..17A}. If the frequency of the
recently discovered high frequency quasi-periodic oscillations found
in IGR J17091--3624 indeed scales with mass, the black-hole mass of
this system should be similar to that of GRS 1915+105 \citep[see][for
a discussion]{2012arXiv1201.2106A} and the source should be at very
large distance (i.e., outside the Galaxy) with a very high quiescent
luminosity.  We expect that if GRS 1915+105 would turn quiescent again
in the future, it should also exhibit a rather high quiescent
luminosity. Together with IGR J17091--3624, GRS 1915+105 would also be
an excellent candidate to test the ADAF explanation for the quiescent
luminosities of black hole transients.

It is possible that we have not observed IGR J17091--3624 truly in its
quiescent state.  During our two {\it XMM-Newton} observations we
might have caught it in an anomalous faint accretion rate regime which
is orders of magnitude lower than when in outburst but which is also
several orders of magnitude higher than during a true quiescent state.
Recently such a sub-luminous accretion state was also possibly
observed for the black hole transient GS 1354--64
\citep[][]{2011ApJ...734L..17R} and several neutron star systems (both
transients as well as persistent sources) have also shown enigmatic
sub-luminous accretion behavior
\citep[e.g.,][]{2002ApJ...579..422W,2005A&A...440..287I,2009A&A...495..547D,2010A&A...524A..69D,2010MNRAS.404.1591D,2011MNRAS.414L.104D}.
The photon index of the observed spectra is consistent with such an
interpretation but the errors are large and as discussed above it is
also fully consistent with the spectra typically observed for
quiescent black hole transients. Our physical understanding of such
long-lived low-level accretion states is still quite limited and is
not easily explained using the disk instability model which has been
used to explain X-ray binary outbursts \citep[see][for a
review]{2001NewAR..45..449L}. The likely variability we have seen
between the two observations might hint at the fact that the source is
not yet quiescent, however, quiescent variability has been observed
for several other black hole transients
\citep[e.g.,][]{2002ApJ...570..277K,2004ApJ...611L.125H}. If the
orbital period of the system is indeed very large, it is plausible
that the source was indeed in quiescence, but if the orbital period
turns out to be very similar to the majority of black hole transients
($<$100 hours), then it is very likely that IGR J17091--3624 was
indeed in a sub-luminous accretion state and that its quiescent state
could be quite fainter than what we have observed during our {\it
XMM-Newton} observations. Clearly, the distance towards the source and
the orbital period of this system need to be determined more
accurately before this source can be put more clearly within the
context of the other black hole transients and with GRS 1915+105 in
particular.

\noindent {\bf Acknowledgements.}\\ RW acknowledges support from a
European Research Council (ERC) starting grant. RW and YJY acknowledge
support from The European Community's Seventh Framework Programme
(FP7/2007-2013) under grant agreement number ITN 215212 Black Hole
Universe. We thank Nathalie Degenaar for giving valuable comments on
an earlier version of this paper. We acknowledge the use of data
obtained through the{\it XMM-Newton} data archive. This research has
made use of NASA's Astrophysics Data System Bibliographic Services.


\label{lastpage}
\end{document}